\documentclass[prl,twocolumn,showpacs,preprintnumbers,amsmath,amssymb]{revtex4}
\newcommand{\ket}[1]{\left | #1 \right\rangle}

\usepackage[dvips]{graphics}
\usepackage{tikz}

\begin{document}

\title{Spin, statistics, spacetime and quantum gravity}

\author{Chiara Marletto and Vlatko Vedral}
\affiliation{Clarendon Laboratory, University of Oxford, Parks Road, Oxford OX1 3PU, United Kingdom and\\Centre for Quantum Technologies, National University of Singapore, 3 Science Drive 2, Singapore 117543 and\\
Department of Physics, National University of Singapore, 2 Science Drive 3, Singapore 117542}

\date{\today}

\begin{abstract}
\noindent We explore the possibility that the connection between spin and statistics in quantum physics is of dynamical origin. We suggest that the gravitational field could provide a fully local mechanism for the phase that arises when fermionic and bosonic particles are exchanged.  
Our results hold even if the symmetry of space and time is Galilean, thus establishing that special relativity is not needed to explain the existence of spin (although it does motivate the introduction of creation and annihilation
of particles, but this is a separate issue). We provide a model for the coupling between a particle of general spin and the gravitational field and discuss it within the context of both the equivalence principle and the Sagnac effect.  This leads us to present a new experiment for testing the quantum nature of the gravitational field. 
\end{abstract}


\maketitle                           

Physicists are usually taught at their mother's knee that Dirac invented a new ingenious way of doing physics. Instead of being guided mainly by experiments, he used higher-level principles (such as the underlying symmetry of the problem) to guess the relevant equations that ought to be satisfied both quantum-mechanically and relativistically. This is how he arrived at his own famous equation, and -- so we are told -- out came the property known as the spin \cite{Dirac}. We are therefore let to believe (as far as almost all textbooks are concerned) that spin is a consequence of special relativity. But, as e shall explain, this is a misconception.
 
Furthermore, and as part of the same physics folklore, Pauli's logic \cite{Pauli} is taught as substantiating the connection between the value of the spin and the resulting particle statistics. Pauli's proof relies on the theory of special relativity, too. This contributes to the generally accepted view that not only is relativity crucial for the existence of spin, but it is also necessary to establish the connection between spin and statistics (we are ignoring the subtleties involving additional assumptions such as the boundedness of energy, which does not affect anything we will conclude in this paper). However, this too is a misconception (most textbooks are, again, guilty as charged). 

We will first dispose of the two aforementioned mistaken beliefs. In particular, we will demonstrate that the necessity for spin is of a much more general kind. In short, it is a consequence of spacetime transformations being linear. Furthermore, we will show that the spin-statistics connection has nothing to do specifically with the Lorentz transformations, but is instead only a consequence of the rotational symmetry of exchanging particles. The quantum phase due to the said exchange of particles, however, cannot be passive. It is not, and cannot be, just a consequence of relabelling particles, for we know that it can be detected by imprinting it into a relative phase of a qubit controlling the exchange (as discussed in \cite{MV} and later in this manuscript). 

This realisation will lead us to look for a dynamical mechanism underlying the phase acquired when swapping fermionic and bosonic systems. Here we propose that such a mechanism is mediated by a field that should couple to all particles to a different degree depending only on the value of their spin, but otherwise completely equally. We conjecture that at present the gravitational field is the only existing candiate for the mediator of particle statistics and propose an experiment to test this idea. This does not, in principle, rule out the possibility that another field different to gravity can also be responsible for the quantal exchange phase; however, the intimate connections of our proposal with the equivalence principle and the Sagnac effect add weight to the conjecture that gravity is by far the current best guess. We conclude that this leads us to a novel way of revealing the quantum nature of gravity. 

{\bf Spin as a consequence of linearity of spacetime transformations.} Let us start with a non-relativistic particle obeying the Schr\"odinger equation. This equation is, of course, satisfied irrespective of whether the particle is a fermion or a boson since there is no spin dependence at this level. Formally, this means that the Schr\"dinger field can be second-quantized (to allow for the creation and annihilation of particles and superpositions of different number of particles) with either commuting or anti-commuting operators without reaching any contradictions. 

However, we will now use Dirac's logic to argue that the space and time need to be treated on an equal footing even in the non-relativistic system. This is because the Galilean transformations are linear as far as the space and time of different frames are concerned (i.e. $t'=t$ and $x'=Rx+vt+d$). In other words, we are looking for an equation with first derivatives in both space and time which implies the Schr\"odiner equation \cite{Leblond}.

The simplest such equation is readily found using the Dirac method of matrices, 
\begin{equation}
(Ai\hbar\frac{\partial}{\partial t} + {\bf B\cdot p} + C)\Phi (x,t) = 0
\end{equation}
where ${\bf B\cdot p} = B_x p_x+B_y p_y+B_zp_z$. We require that $(Ai\hbar\frac{\partial}{\partial t} + {\bf B\cdot p} + C)^2=2m(i\hbar {\partial}/{\partial t} - p^2/2m)$, which is the Schr\"odinger equation, normalised by the multiplying $2m$ factor for convenience. It is straightforward to show that the simplest choice for $A$, $B$s and $C$ are $4$ by $4$ matrices of a specific form \cite{Leblond} whose details do not concern us here (the choice is not unique, as we will see, and different values will correspond to different spins). The main point is that the state $\Phi$ is in this case a $4$ vector composed of two spinors, $\phi$ and $\xi$. They are spinors because they obey the spinor algebra and transform like spinors under rotations (to be discussed more below). 

Therefore, the notion of a spinor arises naturally within the Galilean context too. We obtain a four dimensional description, with both of the spinors satisfying the original Schr\"odiner equation. There is a redundancy since we have that $\phi = -{\bf \sigma \cdot p}/2 \xi$ where ${\bf \sigma p} = \sigma_x p_x + \sigma_y p_y+\sigma_z p_z$ where ${\bf \sigma}$s are the usual Pauli matrices. Therefore, the $\phi$ and $\xi$ spinors are not independent. This, however, naturally encodes spin $S=1/2$ just as in Dirac's case and one could think of the two spinors as corresponding to particles and anti-particles respectively. 

Three observations are now in order that are directly relevant for this paper. 
Firstly, an equation of the above type could be written for any spin by using a trick due to Majorana. Namely, any spin $S$ could be written in the basis of $S$ spin $1/2$ systems whose states are confined to live in the symmetric subspace. For example, a spin one (massive) particle has three orthogonal states that could be written in the basis of $|00\rangle$ ($S_z=-1$), $(|01\rangle + |10\rangle)/\sqrt{2}$ ($S_Z=0$) and $|11\rangle$ ($S_Z=1$) which span the symmetric subspace of two spin half systems. All of these higher spins satisfy the same non-relativistic linear equation above with the appropriate respective forms of the matrices involved.  Spin zero in this case resembles the Feshbach-Villars treatment of the Klein-Gordon equation \cite{Feshbach}.

Secondly, any interaction of the particle with the electromagnetic field could be introduced through the usual ``minimal" coupling $p\rightarrow p-eA$, where $A$ is the electromagnetic vector potential. When this is done, and the resulting spin equation is then squared to obtain back the Schr\"odinger equation, we in fact obtain the Pauli equation with the $1/2 SB$ term through which the spin couples to the external field. This is interesting as the usual way of arriving at Pauli's equation is through the $c\rightarrow \infty$ limit of the Dirac equation. Here instead we have done so by being in the non-relativistic regime since the beginning.

Thirdly, when a Galilean transformation is applied, the spinor transforms as
\begin{equation}  
\phi' (x',t) = e^{i f(x,t)} D^{1/2}(R) \phi (x,t) \; ,
\end{equation}
where $f(x,t) = 1/2mv^2 t + mvRx+C$ and $R$ is the rotation matrix and $D^{1/2}(R)$ is $SU(2)$ spin half representation of rotation. It is this transformation that allows us to call the entities $\phi$ and $\xi$ spinors. Any higher level spin can be handled through tensor product of spin half as mentioned above so that $D^{1}(R) = D^{1/2}(R)\otimes D^{1/2}(R)$ and so on.  We note that the full $2\pi$ rotation around any axis leads to the phase $(-1)^{2S}$. 

We will now need the transformations of the state under spatial rotations only (the full transformation was mentioned for the sake of completeness, but it will not be needed in the rest of the paper). However, because we would like to talk about the connection between spin and particle statistics, we will now second quantise the spinors and think of them as a field. In that case, 
a general spin $S$ field satisfying the second-quantized equations will transform under the general rotation as
\begin{equation}
U(R)\phi (x) U^{\dagger}(R)=D^S(R^{-1}) \phi (Rx)
\end{equation}
which is no different to how the Dirac field transforms \cite{Weinberg}. The transformation is intuitively clear: when the spatial rotation is applied to the state, then the rotation must affect both the orbital as well as the spin part (since the coordinate system has been rotated). In that sense,  the non-relativistic linear equations can be second-quantized without any difficulty and following the same procedure as in the case of special relativity.

{\bf The spin-statistics connection, revisited.} Now that we understand that the rotations are represented in the same way in special and Galilean relativity, the spin-statistics connection can simply be presented as a rotation that exchanges fields at two points $x$ and $y$ (at the same time). Let us for simplicity look at the two point vacuum correlation $\langle \phi (x)\phi (y)\rangle$ \cite{Zumino}: 
\begin{eqnarray}
& & \langle 0|\phi (x) \phi (y) |0\rangle \nonumber \\
 & = &  \langle 0| U^\dagger (R)\phi (x) \phi (y) U(R)|0\rangle \nonumber \\
 & = &  \langle 0| U^\dagger (R)\phi (x)U(R)U^\dagger (R) \phi (y) U(R)U^\dagger (R)U(R)|0\rangle \nonumber \\
& = & (-1)^{2S} \langle 0|\phi (y) \phi x) |0\rangle
\end{eqnarray}
where the first equality follows from the fact that the vacuum is rotationally invariant, the second equality follows from the unitarity of $U(R)$ and the last equality is a consequence of the fact that the full swap is the same as the $2\pi$ rotation. Therefore, integral spin does not gain a phase upon exchange whereas multiples of half integral spin obtain a $\pi$ phase shift as expected. 

We have so far showed that the property of spin naturally arises in the non-relativistic context, and that it leads to the correct connection with statistics simply because the argument only involves behaviour under rotations (which are present in both Galilean as well as Lorentzian spacetimes). We believe that all the results so far are known, although our presentation is original. These results are important because they allow us to discuss experiments that are based on the spin-statistics connection (in the next section) in a non-relativistic setting, without contradictions - a point that had already been elucidated in \cite{BERRY}, with a completely different rationale based on topological phases. 

{\bf A dynamical mechanism to acquire the quantal phase.} Now, however, we would like to argue that the exchange phase is acquired dynamically and locally and then present a mechanism for this. We believe our argument to be novel. It is natural to think that the exchange phase is acquired dynamically for two reasons. First of all, every other quantum phase is provably acquired by local means \cite{MV}. This is to be expected since we believe that quantum physics does obey the locality principle, meaning that actions in one location can only affect the elements of reality pertaining to that location (the elements of reality being the relevant Hermitian operators). Secondly, the phase can be detected by performing a conditional swap of particles and therefore converting the exchange phase into a relative phase between orthogonal qubit states (as we described in \cite{MV}). It is therefore natural to assume that the particle statistics, despite being conventionally encoded with commuting or non-commuting operators in field theory, does - in fact - have a dynamical origin. 

Since the gravitational field couples to all particles the same way, we believe that it is the most natural candidate to implement the exchange phase. Physically, the idea is that exchanging two particles is a rotation, and a rotation, because it is accelerated motion, could be thought of as a particular metric on spacetime. This is because of the equivalence principle, stating that acceleration and gravity cannot be locally discriminated. In fact, in analogy with the electromagnetic field, the gravitational potential in the frame rotating uniformly at the rate $\Omega$ is ${\bf A}_g= \Omega {\bf r}/c$ where ${\bf r}$ is the radius of rotation and $c$ is the speed of light. This leads to the magnetic like component of the gravitational field ${\bf B}_g = \nabla \wedge {\bf A}_g = 2{\bf \Omega}$. It is this field that couples to the spin if we assume the minimal coupling as in the case of the electromagnetic field. 

This leads us to the following Pauli-like equation
\begin{equation} \label{PAULI}
(\frac{p^2}{2m} + \hbar {\bf S \cdot \Omega}) |\psi\rangle = i\hbar \frac{\partial}{\partial t} |\psi\rangle
\end{equation}
describing how the spin of a rotating particle couples to the spacetime rotating at rate $\Omega$. The phase upon swap is $e^{i S\Omega t} = e^{i S\Omega 2\pi/\Omega}=(-1)^{2S}$. We therefore obtain the fermionic and bosonic exchange phase shift. We note that a different speculation leading to conjecture the connection between gravity and spin was presented in \cite{Unnikrishnan}. 

We assumed a uniform rotation to swap particles at $x$ and $y$ only for simplicity.  A time dependent frequency $\Omega (t)$ would still give the same shift when the full swap is performed (because $\int_0^{2\pi} (d\omega/dt) dt  = 2\pi$). 

{\bf Experimental implications.} Can our theory be tested? One testable aspect has to do with the locality of the exchange phase, as already mentioned in \cite{MV}. Namely, if instead of performing the full swap of two particles, we execute a partial swap, our theory predicts that the phase will be equal to the corresponding fraction of the full phase. This can certainly be tested with the current technology.  

In addition, our mechanism and the standard quantum field theory description in terms of fermionic algebra make different predictions about states with an odd number of fermions. In principle, this difference could be detected, following these steps. 
Note first that the equation \eqref{PAULI} implies that by moving a single particle with half-integer spin along a circle, for example an electron ($S=\frac{1}{2}$), one would have a $-1$ phase appear to multiply its original quantum state. Specifically, calling $\ket{\psi_{0}}$ the state where the electron is in a given position, dragging it along a circle back to its original position, would lead to this phase:
\begin{equation}
U(R)\ket{\psi_{0}}=(-1)^{2S}\ket{\psi_{0}}
\end{equation}
where $U(R)$ is the full rotation encountered earlier (see figure 1a). 

On the other hand, considering the standard fermionic algebra approach a single fermionic mode moved about in a circle would not acquire a phase, contrary to what \eqref{PAULI} predicts.  

The phase $(-1)^{2S}$  is impossible to observe by making measurements on the spin-$\frac{1}{2}$ particle only (since it is then only a global phase). However, it is possible to adapt the scheme from \cite{MV} in order to detect the difference between the two cases. One can use a controlled $U(R)$ gate, which is controlled by a qubit initially prepared in a superposition of its computational basis states. This qubit controls, coherently, whether the $U(R)$ operation is applied, or not. If the qubit is in the state $|0\rangle$, then no swap occurs, while if the qubit is in the state $|1\rangle$, the transformation $U(R)$ occurs. Using equation \eqref{PAULI}, we expect the fermionic phase to be transferred to the qubit, so it can be detected as the phase between the qubit states $|0\rangle$ and $|1\rangle$. In the case of the standard quantum field theory prediction, we have instead no extra phase on the control qubit. Performing this experiment, by interfering the control qubit at the end, leads one to testing our proposed mechanism of generation of the phase. 

\begin{figure}[htb]
\includegraphics[width=\columnwidth]{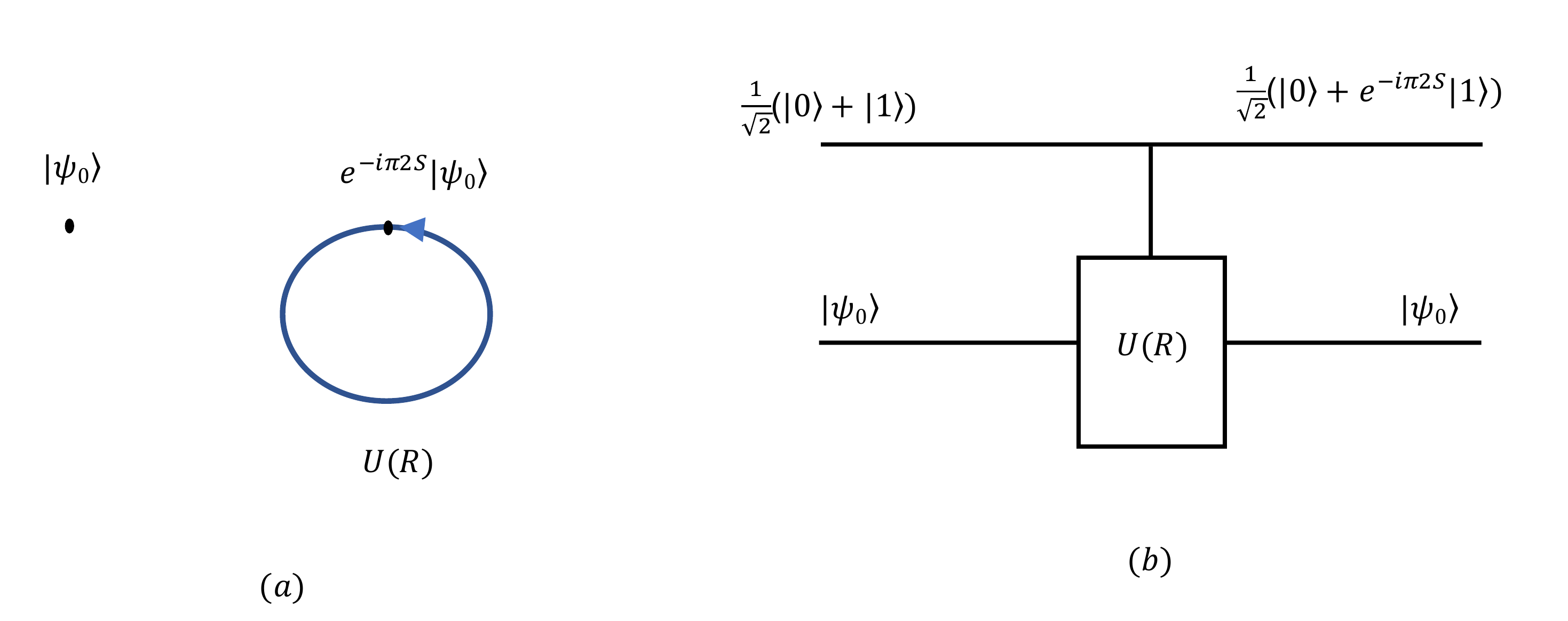}
\caption{Dynamical acquisition of spin-statistics phase. (a): Prediction of equation \eqref{PAULI} for the full swap operation $U(R)$; (b): proposed experiment to test this prediction when $S$ is half-integer, with a controlled-$U(R)$ operation.}
\end{figure}

It is also interesting to connect our dynamical, local theory of phase acquisition with the recently proposed experiments to test non-classicality in the gravitational field, \cite{MAVE, SOUG}. The interaction described by equation \eqref{PAULI} is local, because it happens at the location where the fermion (or boson) is. So it can be extended to describing the interaction of two particles with half-integer spin $i=1,2$ with $\Omega_i$ being the spacetime geometry at location $i$, as follows:
\begin{equation}\label{PAULI2}
(\sum_{i=1}^{2}\frac{p_i^2}{2m}  + \hbar {\bf S_i \cdot \Omega_i}) |\psi_{12}\rangle = i\hbar \frac{\partial}{\partial t} |\psi_{12}\rangle
\end{equation}
This equation would allow one to predict the exchange phase of say two electrons $1$ and $2$, each with spin $S=\frac{1}{2}$ and each in a superposition of two different locations, as generated dynamically by local interaction with the spacetime geometry, represented here by the field $\Omega$. 
This prediction is identical to that of a second-quantized model with a Hamiltonian that couples the quantised field $\Omega$ (bosonic) with the fermionic modes in four possible locations (a generalisation of the linear quantum gravity equation as presented in \cite{MAVE}). 

\begin{figure}[htb]
\includegraphics[width=\columnwidth]{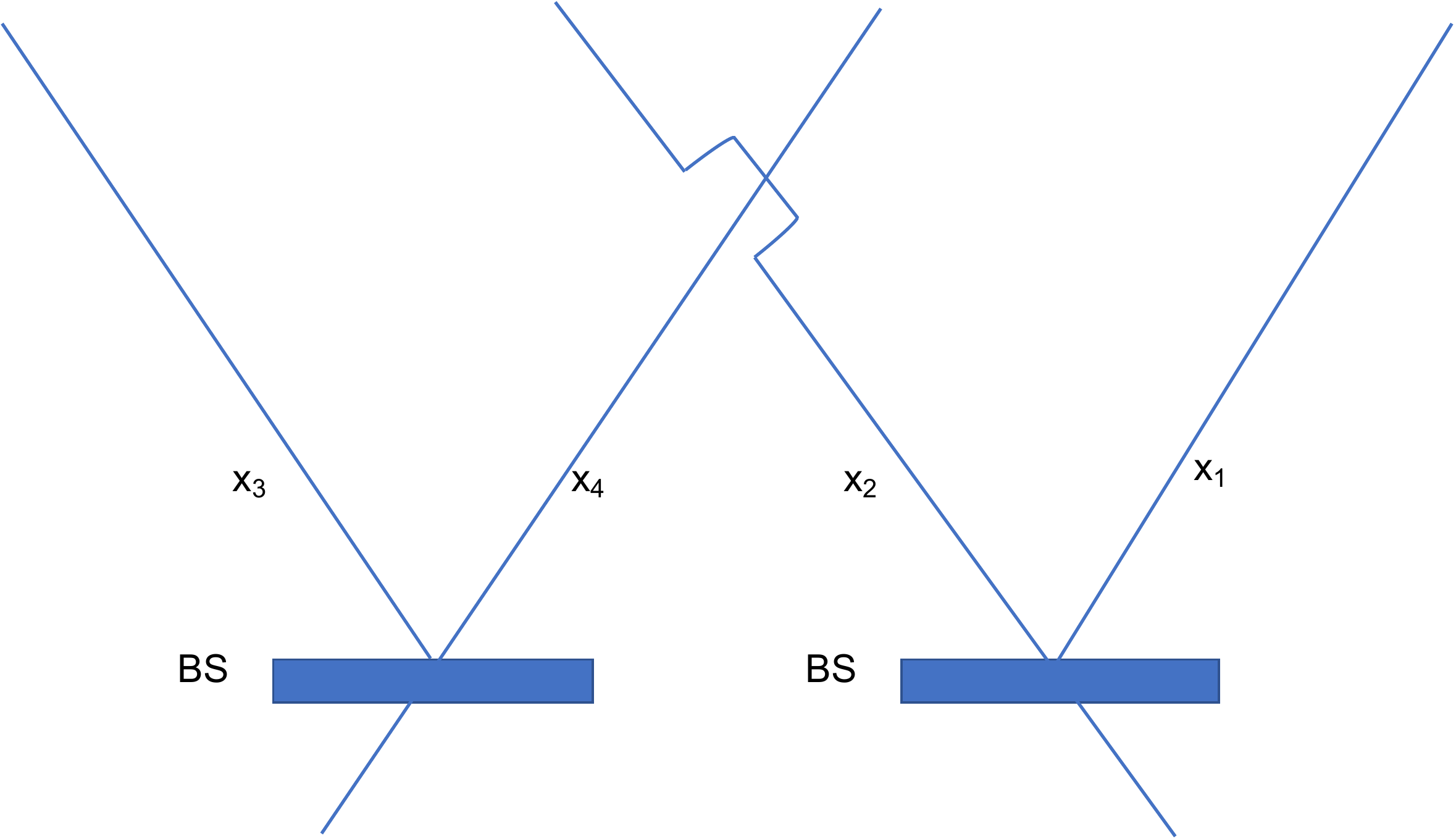}
\caption{Scheme to generate entanglement via swapping the paths of two half-integer spin particles. After going through a beam-splitter (BS) the paths of two electrons ($x_2$ and $x_4$) get swapped, generating entanglement.}
\end{figure}

The phase generated by the interaction has the effect of entangling the two electrons in a much stronger way than theier gravitational interaction does (hence, for simplicity we can ignore the latter in this discussion). The degree of entanglement is a function of the gravitationally mediated phase, which therefore becomes a witness of non-classicality of gravity itself. Being local, it satisfies the criteria set out in \cite{MAVE, MAVE2} for being a good witness of the non-classicality of the gravitational field, should we be able to use it to generate detectable entanglement. A thought experiment would go as follows (see figure 2): first, let two electrons each through a beam splitter. The two electrons (say 1 and 2) are now in the state:
\begin{equation}
\frac{1}{\sqrt{2}}(\ket{x_1}+\ket{x_2})_1(\ket{x_3}+\ket{x_4})_2 \; ,
\end{equation}
where each of the branches $x_i$ indicates a location. Assuming to apply a swap between locations $x_2$ and $x_4$, we obtain:
\begin{equation}
\frac{1}{\sqrt{2}}\left(\ket{x_1}_1\ket{x_3}_2+\ket{x_2}_1\ket{x_3}_2+\ket{x_1}_1\ket{x_4}_2-\ket{x_2}_1\ket{x_4}_2\right)
\end{equation}
Hence this operation leads to a $-1$ phase in one of the branches - thus creating maximal entanglement, which can then be detected by following the usual procedures. Considering partial swaps with varying strenght will lead to intermediate degrees of entanglement, from zero to maximal. Hence a very interesting consequence of our model is that the detection of spin-entanglement in this setup is also an indirect evidence of the quantisation of gravity, if we take equation \eqref{PAULI} to be correct. This experiment is much more within reach than the mass-based ones, which makes this speculation tantalising. 

But can the particular mechanism involving spin-gravity coupling be tested, and what would its implications be? Here the question is harder to answer, because a natural experimental effects to look for are any deviations from the ${\bf S\cdot B_g}$ coupling. One way to do so is to look at the corrections to this coupling term. In the electromagnetic case there would be a term of the form ${\bf S \cdot E\wedge p}$. This term would be responsible for the Thomas precession in atomic physics, and the question is whether there is a gravitational analogue of Thomas precession. The physical explanation would be via a time-dependent gravitational vector potential ${\bf A}_g (t)$ which would give rise to the gravitational field ${\bf E}_g = - \partial/\partial t {\bf A})g(t)$. 

Another interesting conjecture is that one could seek an experiment to tell the difference between coupling with quasi-Newtonian gravity and the GR corrections. We leave the exploration of this interesting point for a future paper. 

\textit{Acknowledgments}: CM thanks the John Templeton Foundation and the Eutopia Foundation. VV's research is supported by the National Research Foundation and the Ministry of Education in Singapore and administered by Centre for Quantum Technologies, National University of Singapore. This publication was made possible through the support of the ID 61466 grant from the John Templeton Foundation, as part of the Quantum Information Structure of Spacetime Project (qiss.fr). The opinions expressed in this publication are those of the authors and do not necessarily reflect the views of the John Templeton Foundation.

\end{document}